\documentclass[12pt,preprint]{aastex}
\usepackage{amsmath,amstext}
\def\teff{$T_{\rm eff}$}
\def\lteff{$\log T_{\rm eff}$}
\def\logg{$\log g$}
\def\smax{{$s_{\rm ad}$}}
\def\sad{{$s_{\rm ad}$}}

\shorttitle{The entropy {adiabat}}
\shortauthors{Tanner et al.}

\usepackage{amsmath}

\begin{document}

\title{{Entropy in adiabatic regions} of convection simulations}
\author{Joel D. Tanner,  Sarbani Basu and Pierre Demarque}
\affil{Department of Astronomy, \\ P.O. Box 8101, Yale University, New Haven CT 06520-8101}
\email{sarbani.basu@yale.edu}

\begin{abstract}
One of the largest sources of uncertainty in stellar models is caused
by the treatment of convection in stellar envelopes. {One dimensional
stellar models often make use of the mixing length or equivalent
approximations to describe convection, all of which depend on various
free parameters.}  There have been attempts to rectify this by using
3D radiative-hydrodynamic simulations of stellar convection, and in
trying to extract an equivalent mixing length from the simulations. In
this paper we show that the {entropy of the deeper, adiabatic layers} in these
simulations can be expressed as a simple function  of $\log g$ and
$\log T_{\rm eff}$ which holds potential for calibrating stellar
models in a simple {and more general} manner.
\end{abstract}


\section{Introduction}

The treatment of convection in stellar envelopes is one of the largest
sources of uncertainty in the interior modeling of late-type stars.
Convection in stellar models is usually described by the mixing length
theory (MLT; B\"{o}hm-Vitense 1958), which represents convection with
a single characteristic length that is proportional to the local
pressure scale height $l=\alpha H_p$, where $\alpha$ is a free
parameter. There are other 1D formulations (e.g., Arnett et al. 2010),
but these are not devoid of free parameters either. MLT and other
formulations define the thermal stratification of the convective
envelope, which is essentially adiabatic, and the primary weakness
affecting these formulations  is that the existence of the freely
adjustable scale factors, like $\alpha$, permits a wide range of
adiabatic structures.

The mixing length parameter is usually held at a constant value for
{stars at all phases of evolution}. Most frequently, this value is
the one needed to model the Sun such that it has the correct radius
and luminosity at the solar age. An issue with the mixing length
parameter is that it is not unique, even for calibrated solar models.
While calibrated models all have the correct radius, even with
chemical composition constrained, the value of $\alpha$ depends on the
atmospheric model (the $T$-$\tau$ relation) used, the equation of
state, and {also} on whether or not diffusion and gravitational
settling of helium and heavy elements are included in the models.
Clearly, $\alpha$ alone does not contain intrinsic information about
convective dynamics, and a value that is suitable for one model may
not be appropriate for another.

The mixing length parameter determines the radius of a stellar model,
and hence predictions of stellar radii can be incorrect. Additionally,
since the parameter is usually held constant in stellar model
calculations, the dependence of convection on the properties of stars,
such as surface gravity, effective temperature and metallicity are
eliminated. This is the case despite the fact that data suggest that
the mixing length parameter should depend on stellar properties such
as metallicity (e.g. Bonaca et al. 2012) and other properties
(Y{\i}ld{\i}z 2006; Lebreton et al. 2001). The limitations of the
mixing length approximation have led to studies of stellar convection
using three-dimensional radiative hydrodynamic (3D RHD) simulations.
Simulations have been applied to dwarf stars (e.g., Ram\'{i}rez et al.
2009), giants (e.g., Ludwig \&  Ku{\u{c}}inskas 2012), and several
targeted studies of individual stars (e.g., Robinson et al. 2004,
2005; Straka et al. 2006, 2007; Ludwig et al. 2009; Behara et al.
2010).

Efforts to systematically study the variation of stellar convection
have been carried out by Ludwig et al. (1995, 1998, 1999), Freytag et
al. (1999), Trampedach \& Stein (2011), Tanner et al. (2013a,b), Magic
et al. (2013), Trampedach et al. (2013). The current focus of research
in the community is to determine how the properties of convection from
3D simulations can be applied to 1D models of stars. {For example,
one of the properties that can be extracted from 3D simulations of
convection is the $T$-$\tau$ relation, which can be used as the outer
boundary condition of 1D stellar models.} Tanner et al. (2014) have
shown that the $T$-$\tau$ relation depends on the properties of a star
and is generally quite different from the approximate models, such as
the Eddington  $T$-$\tau$ approximation (see e.g. Mihalas 1978), or
even semi-empirical ones, such as the Krishna Swamy relation (1966) or
the VAL models (Vernazza et al. 1981). Trampedach et al. (2014) have
made available a suite of $T$-$\tau$ relations derived from 3D
simulations, and codes to use them easily. The second, crucial,
parameter that is the focus of research is the mixing length parameter
itself. 

The idea of determining an effective mixing length parameter from
simulations of convection is not new. Early efforts to derive a
relationship between $\alpha$ and stellar parameters include the use
of 2D simulations by Ludwig et al. (1999) to map out the envelope
specific entropy in the $\log g$-$\log T_{\rm eff}$ plane, and
translate it into a mixing length parameter. However, this was not
widely adopted. More recently, Trampedach et al. (2014) calibrated the
mixing length parameter by matching averages of 3D simulations to 1D
stellar envelope models. They found that this led to $\alpha$ varying
from 1.6 for the warmest dwarf, which is just cool enough to have a
convective envelope, and up to 2.05 for the coolest dwarf in their
grid. Magic et al. (2015) used a different approach and used the
entropy profiles to determine values of the mixing length, from this
they provide a functional form for the mixing length that depends on
\logg, \lteff, and metallicity.  


{In this letter, we use the simulations from Tanner et al.
(2013a,b; 2014), as well as published results of Magic et al. (2013)
and Trampedach et al. (2013) to show that the entropy in the
adiabatic regions of 3D
simulations can be expressed more conveniently in a single-valued
functional form when projected on a rotated \logg-\lteff\ plane.  The
method proposed in this letter builds upon the pioneering work of
others, but offers a few advantages.  First, a single-valued
functional form is convenient from a modelling perspective.  For
example, in stellar evolution codes, the desired stellar model entropy
can be evaluated as the model evolves without the need for
multidimensional interpolation in the \logg-\lteff\ plane.  Second,
and more importantly, calibrating against thermodynamic quantities is
not dependent on particular modelling codes.  In the absence of an
improved model that accurately describes convective dynamics in stars,
the most direct route to improving stellar models through calibration
may be to leverage existing parameterized convection models such as
MLT.  While thermodynamic quantities (in this case the entropy
adiabat, \sad) can always be related to parameters like the mixing
length, the translation renders the calibration model-dependent.  This
is indeed useful if one wishes to calibrate models with a particular
stellar evolution code, but it cannot be applied generally since the
interpretation of parameters such as $\alpha$ is specific to the
model.  Instead, we choose to look at how fundamental physical
quantities, such as the specific entropy, vary in the \logg-\lteff
plane.}

\section{Mixing Length Theory and Convection Zone Entropy}

One of the major weakness affecting models constructed using the MLT
is the freely adjustable scale factor $\alpha$,  which permits a wide
range of adiabatic structures.  {This, and three other free
parameters \citep[see e.g.,][]{hgl99,arnett2010} in the MLT formalism
to account for geometric properties of convection,} set the entropy
profile below the photosphere, and determine the asymptotic limit of
the entropy {(or \sad)} that is reached when convection is
efficient, and the stratification is very near to adiabatic. This is
in turn reflected in a large uncertainty in the calculated radii.

\begin{figure}
\epsscale{0.6}
\plotone{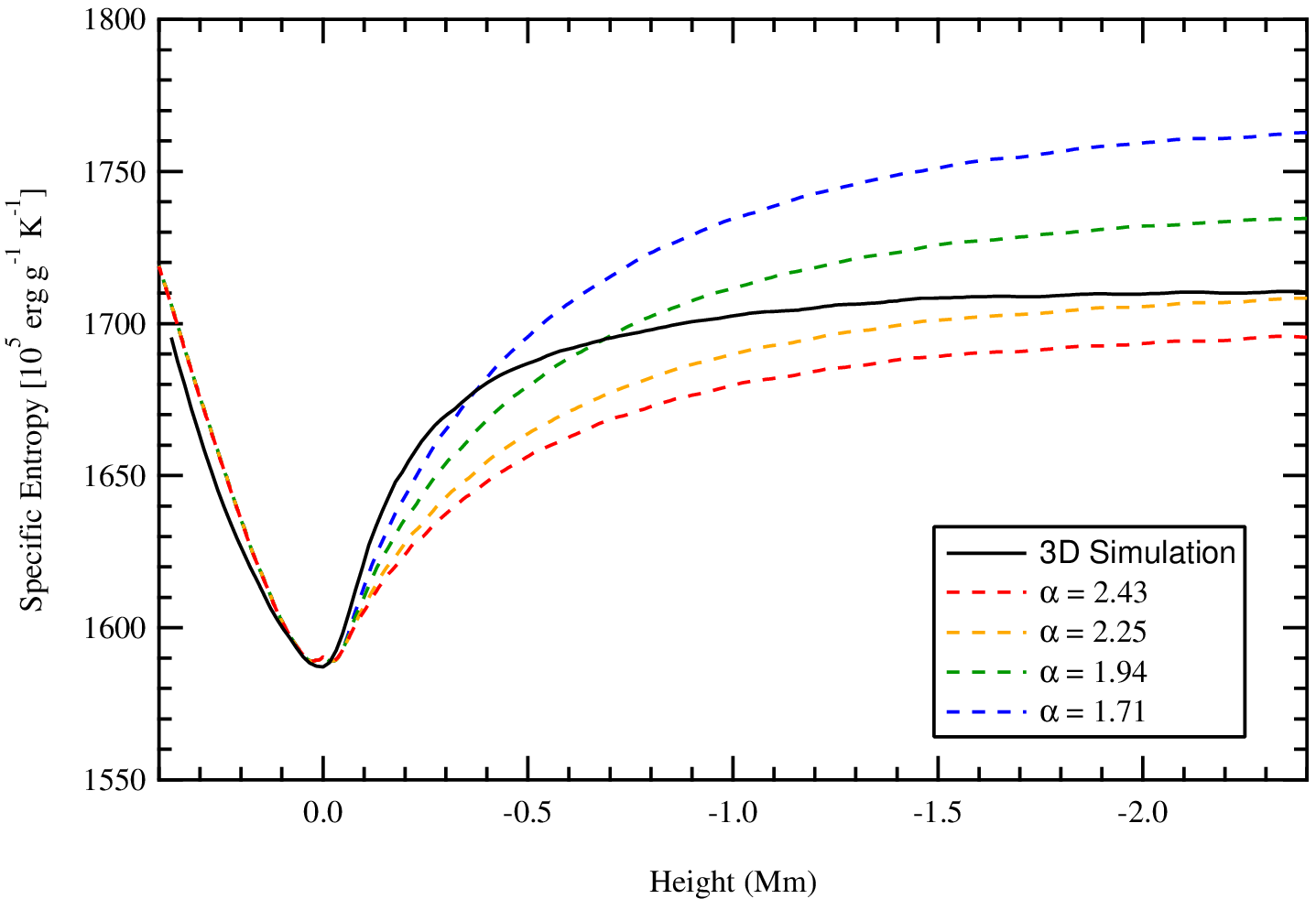}
\plotone{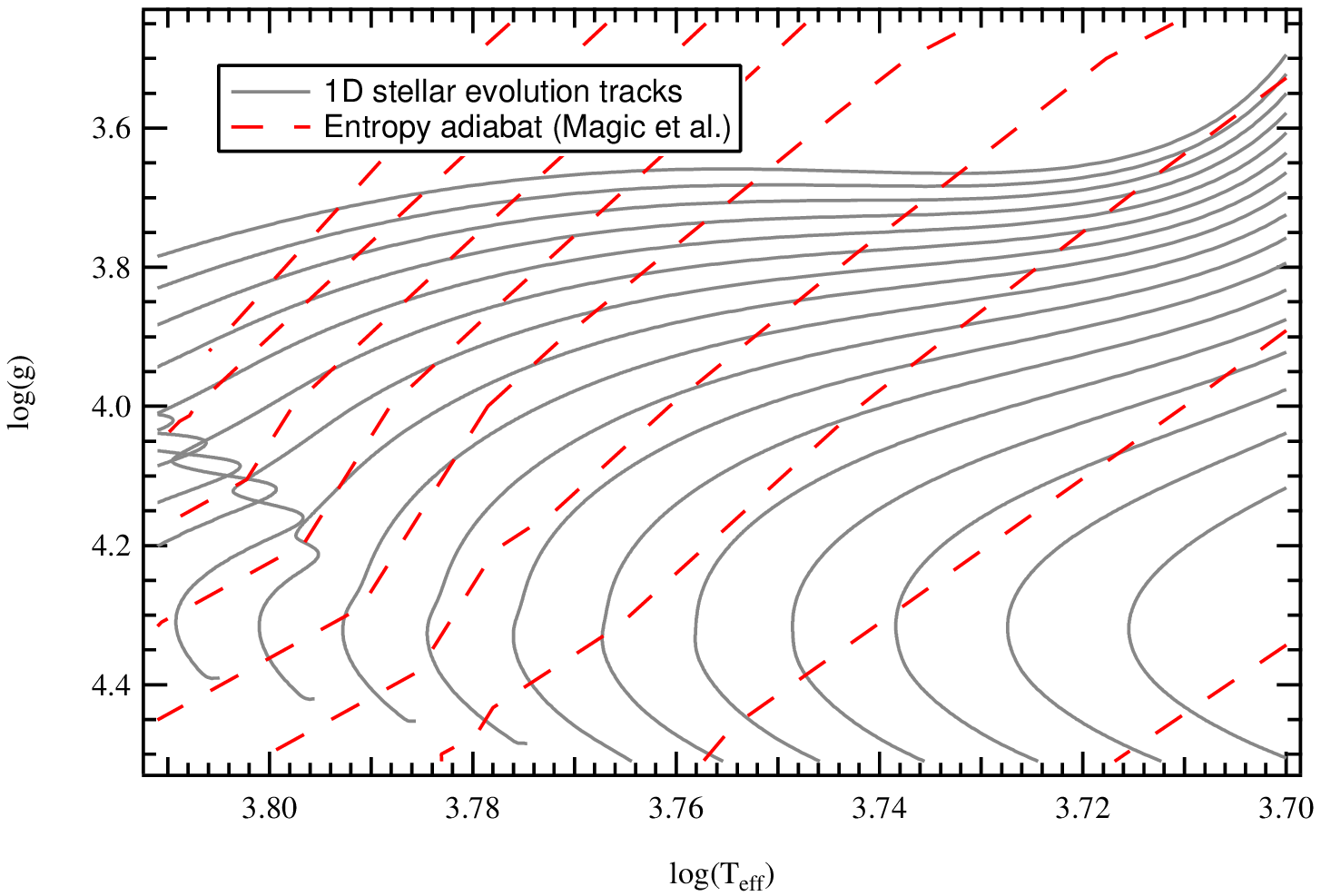}
\caption{{\bf Top panel:} Specific entropy near the surface of several
1D stellar models and one 3D RHD simulation. The models and simulation
all share the same surface parameters {($\log g=4.30$, and $\log
T_{\rm eff}=3.76$)} and chemical composition {($Z=0.001$,
$X=0.754$)}, but the 1D models are computed with different mixing
length parameters, {and so have different envelope entropy (\smax).
The simulation does not contain a mixing length parameter, so the
specific entropy is determined self-consistently and uniquely.
{\bf Bottom panel:} Contours of adiabatic entropy (\smax) in the
\logg-\lteff\  plane, as determined by the 3D simulations {of Magic
\textit{et al.}. The contours of {\sad} (ranging from $16 \cdot 10^8$
to $24 \cdot10^8$ $\mbox{erg}$ $\mbox{s}^{-1}$ $\mbox{K}^{-1}$ from
the lower-left to upper-right) are} equivalent to contours of constant
polytropic K (see e.g. Kippenhahn \& Weigert 1990) and denote the
convective envelope adiabats. Evolutionary tracks for a range of
stellar masses ($M_\odot = 0.75$ - $1.40$), all with the same
composition parameters, and with the same constant value of the mixing
length parameter $\alpha$, are shown for reference}}
\label{fig:mlt}
\end{figure}

With MLT models alone, there is no way to determine which asymptotic
entropy, or adiabat, is correct. To illustrate this, in
Figure~\ref{fig:mlt} we show the specific entropy profiles of four
{1D} stellar models with identical {stellar atmosphere
parameters}, each computed with a different value of $\alpha$.
{The specific entropy in both 1D models and 3D simulations was
calculated with the OPAL \citep{2002ApJ...576.1064R} equation of state
tables.}  Near the surface there exists a {steep entropy} gradient
where radiative transfer of energy dominates, and the stratification
is convectively stable. Further down, the entropy reaches a minimum
and the entropy gradient switches sign, indicating that the region is
convectively unstable.  The entropy gradient continues to flatten with
depth, with the entropy approaching a near-constant value {\smax} that
depends on $\alpha$, and remains roughly constant {throughout the
convective region until the effect of overshoot near the interior edge
of the convective envelope changes the profile again.} 

{One advantage of 3D simulations over 1D models is that simulations do
not have an arbitrarily set mixing length parameter, and instead
converge to a thermal structure that self-consistently links the deep
adiabatic layers to the radiative atmosphere.} Also shown in
{the upper panel of}
Figure~\ref{fig:mlt} {is the mean entropy profile for a 3D
simulation with the same \logg\ and \lteff\ as the 1D models.} There are
no free parameters (beyond factors for artificial viscosity and the
subgrid scale model), so the resulting entropy profile is unique to
the surface gravity, effective temperature and chemical composition of
the simulation. Comparing the simulated entropy profile to the MLT
models, we see that there is a value of $\alpha$ that can reproduce
the simulated \smax. However, the {{complete entropy profile}} in the
simulation cannot be matched by any of the MLT models, and this can be
for a number of reasons, such as the use of an inconsistent $T$-$\tau$
relation or more likely, the absence of dynamical effects in the 1D
models. We shall concentrate only on {\smax} in our approach to mixing
length calibration.  This is similar to the recent work of Magic et
al. (2015), where the entropy adiabat is related to the mixing
length parameter; what we show here, is that the {{evolution of}}
\sad\ {{could potentially be described as a function of a single
variable, which would be simpler to implement in 1D stellar evolution
codes.}}

\section{The entropy calibration}

\begin{figure}
\epsscale{0.7}
\plotone{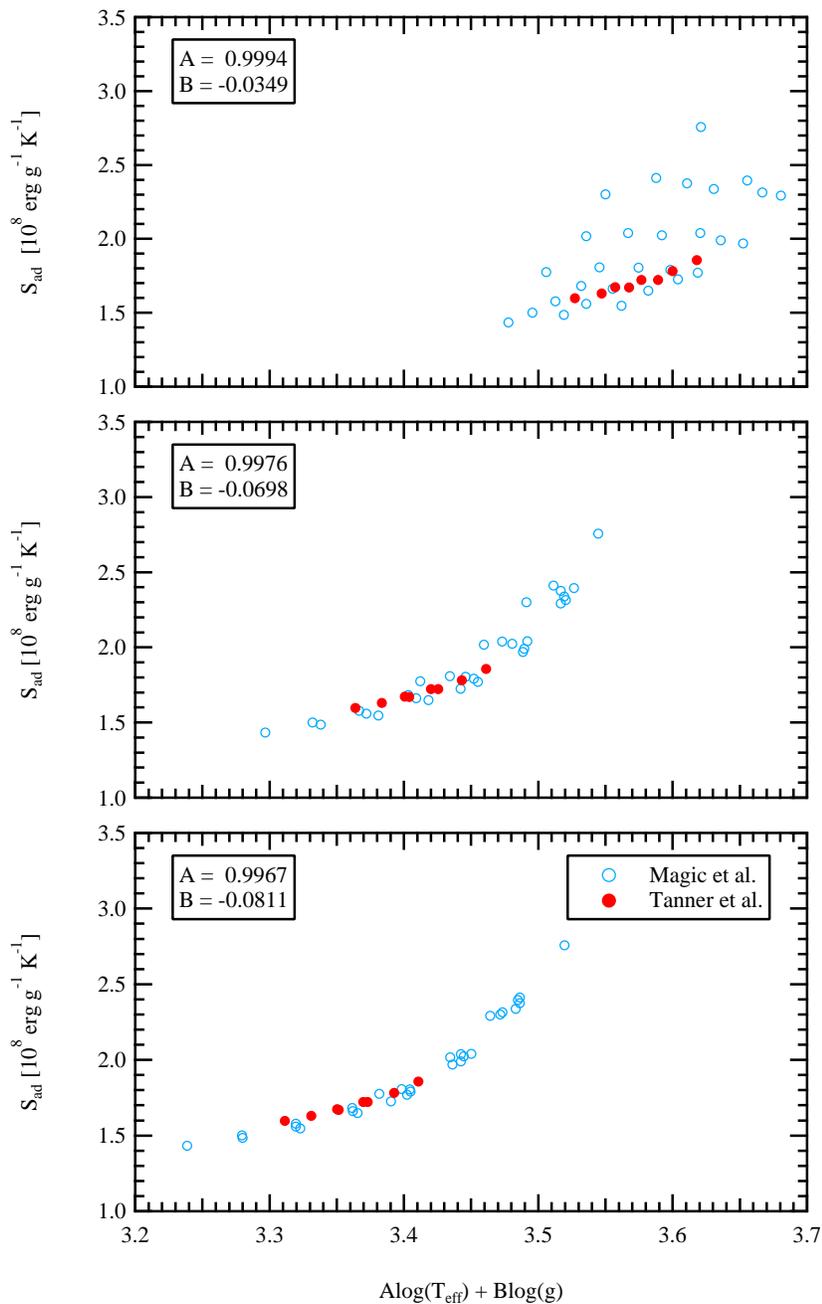}
\caption{{Adiabatic entropy (\sad)} from 3D simulations {of a
particular chemical composition}, presented along different
projections of the \logg-\lteff\ plane. Variation in simulated entropy
follows a single value functional form {when presented against} a
particular projection that is aligned with the adiabats. The lower
panel is the projection that shows the least scatter in a regression
model, which is the projection that is aligned with the adiabats
{(i.e. the contours {in the lower panel of Figure
\ref{fig:mlt}}).}}
\label{fig:rotate}
\end{figure}

In the lower panel of Figure~\ref{fig:mlt}, we show contours of
constant \smax\ as obtained from 3D simulations by Magic et al. (2015)
plotted on the \logg-\lteff\ plane. Also shown on the plot for
reference are evolutionary tracks {(computed with the Grevesse and
Sauval (1998) mixture and metallicity $Z=0.018$), which are included
to show the region of main-sequence stellar evolution.}   {One
striking feature of the \smax\ contours is that they are nearly
parallel, and for a particular chemical composition, \smax\ appears to
be a smooth function of \logg\ and \lteff .  The smoothly-varying
nature of the \smax\ contours leads us to believe that a simple
{projection} of the \logg-\lteff\ plane may sufficient to exploit
the fundamental relationship between \sad, \logg, and \lteff.  We show
that this is indeed possible in Figure~\ref{fig:rotate}, where
simulations performed independently by Tanner et al. (2013a,b; 2014)
and Magic et al. (2013) are presented on different {projections} of
the \logg-\lteff\ plane.} These simulations were performed with
different codes and with different radiative transfer schemes, and
while the simulations had similar equations of state and
metallicities, they differed in their atmospheric structures: the
Tanner et al. simulations assume gray atmospheres  while Magic et al.
do not. For all these simulations, the envelope entropy, {\smax}, can
be projected on to a one-dimensional curve when plotted against a
linear combination of \lteff\ and \logg\, {(i.e. the \logg-\lteff\
plane  becomes $A\log T_{\rm eff}+ B\log g$).}

\begin{figure}
\epsscale{0.8}
\plotone{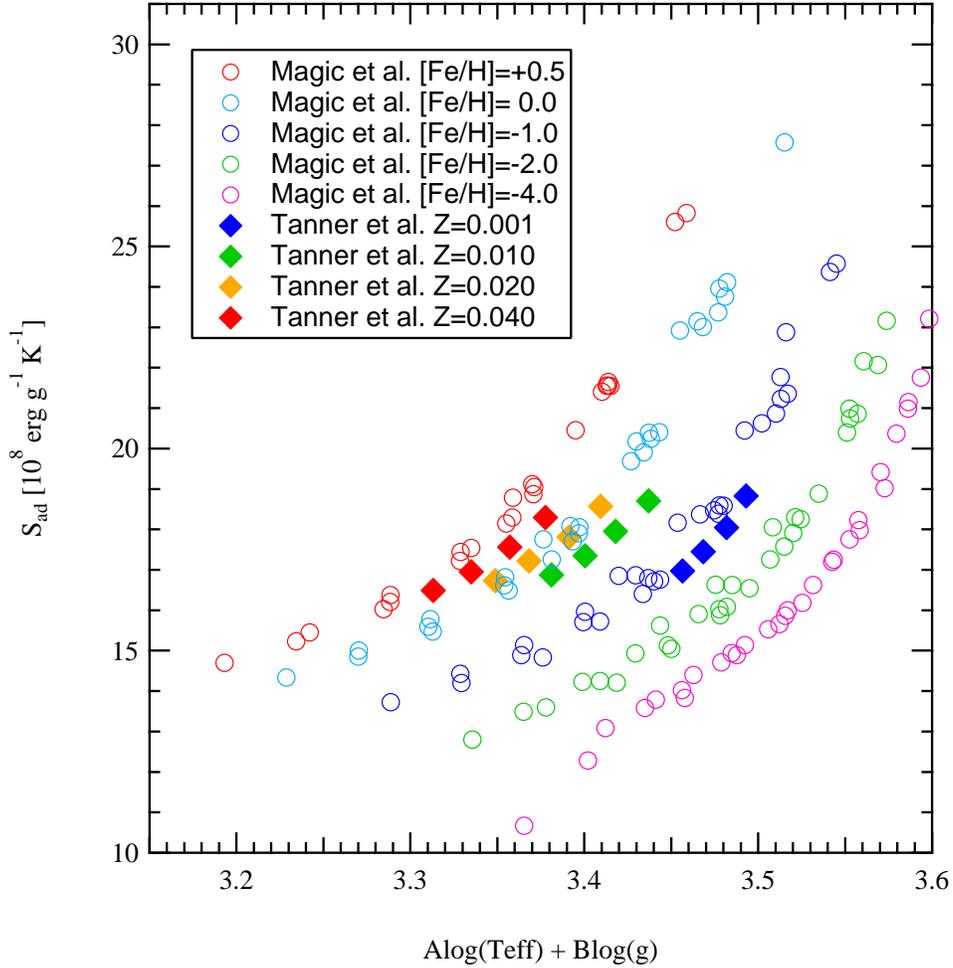}
\caption{{{Adiabatic entropy} from sets of 3D simulations with varied
chemical compositions (similar to lowermost panel in
Figure~\ref{fig:rotate}) presented along projections in the
\logg-\lteff\ plane. Each metallicity requires different coefficients
for $A$ and $B$, which correspond to different adiabatic contours in
the \logg-\lteff\ plane.}}
\label{fig:metal}
\end{figure}

{In this work, the precise values of the constants $A$ and $B$ for
each metallicity were selected with a non-linear least squares
minimization to a pre-determined function.   First, a function of the
form $(s_{\rm ad}-s_0) = \beta \exp\left( (x-x_0)/\tau \right);
x=A\log T_{\rm eff}+ B\log g$ was selected by visual inspection to
represent the dimensionally-reduced dataset.  The choice of function
is arbitrary, but the authors note that the resulting parameters $A$
and $B$ are not particularly sensitive to this, provided that the
function can adequately reproduce the variation of \sad.  This
function function comprises six parameters that define the
relationship of \sad\ across the \logg-\lteff\ plane, and the
least-squares minimization algorithm of \citet{2009ASPC..411..251M}
was then used to determine their values (listed in Table
\ref{table:params}).  The process is repeated for different convective
envelope compositions (see Figure~\ref{fig:metal}), each of which
require a unique projection of the \logg-\lteff\ plane. This fitting
process effectively reduces the dimensionality of the initial
variation of \sad\ by projecting the \logg-\lteff\  plane onto an axis
that is aligned with the convective envelope adiabats.  The process
used in this work is only one possible method for reducing the
dimensionality of the problem, and further study using other
statistical tools, such as principle components analysis, may yield
additional insights into the fundamental relationship between
convection zone entropy and the stellar surface parameters.}

\begin{table}
\begin{center}
\begin{tabular}{ |c|c|c|c|c|c|c| } 
 \hline
 [Fe/H] & $A$ & $B$ & $s_0$ & $x_0$ & $\beta$ & $\tau$ \\
 \hline
 0.5     &  0.9961   & -0.0884  &   1.396    &   3.435   &    0.929   &   0.1009 \\
 0.0     &  0.9967   & -0.0811  &   1.336    &   3.485   &    1.051   &   0.1056 \\
-1.0     &  0.9974   & -0.0720  &   1.304    &   3.540   &    1.127   &   0.0973 \\
-2.0     &  0.9981   & -0.0623  &   1.254    &   3.603   &    1.439   &   0.0899 \\
-4.0     &  0.9985   & -0.0553  &   1.104    &   3.606   &    1.216   &   0.0985 \\
\hline
\end{tabular}
\caption{Parameters for the function fit to \sad\ from Magic
\textit{et al.} simulations in the projected \logg-\lteff\ plane
($A\log T_{\rm eff}+ B\log g$).}
\label{table:params}
\end{center}
\end{table}

Since the convection zone {adiabatic entropy value} in a stellar
model is determined by the mixing length parameter, the curves in
Figure~\ref{fig:metal} basically show how we need to change $\alpha$
as a function of \logg\ and {\lteff}. Of course, the first step would
be to determine which numerical value of $\alpha$ yields a particular
\smax\,  given the rest of the physics to set the mixing length scale,
and thus determine how \smax\ changes with $\alpha$. {After setting
the relationship between \smax\ and $\alpha$, all that is required is
to follow {this} relationship (i.e. the curve in
Figure~\ref{fig:metal})} as the star evolves. Since each time step in
a stellar evolution calculation changes \logg\ and \teff\, we will
need to keep changing $\alpha$ as we evolve a model. 

\begin{figure}
\epsscale{1.0}
\plotone{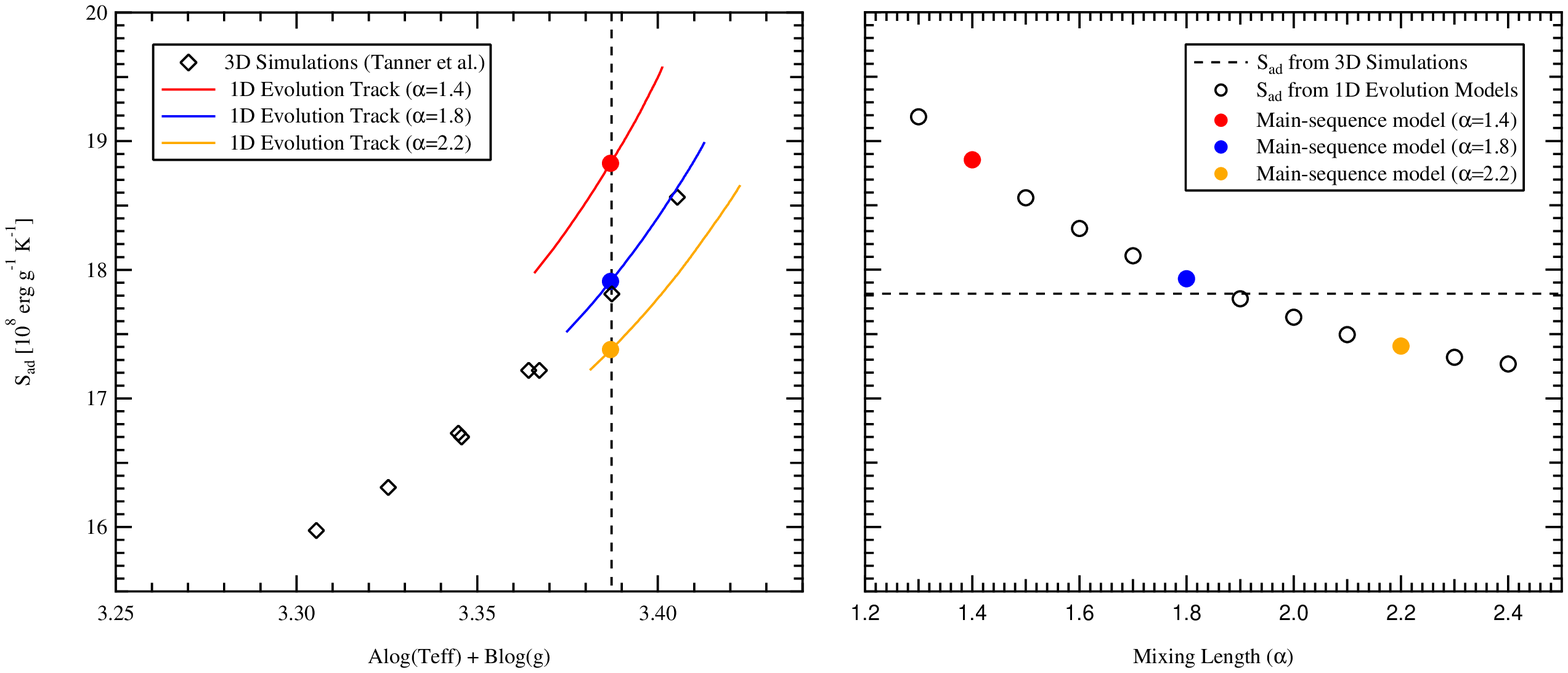}
\caption{{Stellar envelope specific entropy \smax\  from 3D RHD
simulations and 1D MLT stellar models. \emph{Left}:  Simulations and
evolution tracks have the same convective envelope chemical
composition, and \smax\ has been plotted in a manner similar to Figure
\ref{fig:metal}}.  The three evolution tracks {are  $1.1
\mbox{M}_\odot$ models evolved to approximately $8.6\; \mbox{Gyr}$},
and were constructed with different values of the mixing length
parameter, $\alpha$. \emph{Right}: variation of {\smax} in 1D
{main-sequence} stellar models as a function of ${\alpha}$.  The
stellar models have the same surface conditions, but only one value of
$\alpha$ produces a stellar model matching the specific entropy from a
3D simulation (dashed line) with the same surface conditions.  Note
that this calibration mapping of $\alpha$ into \smax\ depends on the
choice of the $T$-$\tau$ relation.}
\label{fig:entropycal}
\end{figure}

The two panels of Figure~\ref{fig:entropycal} outline in principle the
steps that must be taken to translate the {adiabatic} specific
entropy derived from a 3D simulation into the corresponding entropy
calibrated value of $\alpha$ to be used in the 1D MLT stellar model.
Figure~\ref{fig:entropycal}(left) shows, in the same entropy
calibration plane as Figure~\ref{fig:metal}, the locus of a set of 3D
simulations, all with the same {chemical composition}, but with
different values of \smax\  in the deep part.  Also in the left panel,
are three 1D MLT models, all with the same metallicity and surface
conditions as the 3D simulations.  {Presented relative to the
projected {\logg} and {\lteff} coordinates in this way, the evolution
tracks begin on the zero-age main sequence with a relatively low
{\smax}, which increases as the model approaches the terminal age.
For the purpose of demonstrating our calibration method, we will focus
only on the main-sequence phase of the evolution tracks.}  Because in
the 1D MLT models, for a given composition, we have $s_{\rm max} =
f(g, T_{\rm eff}, \alpha)$, all three models were chosen, for the sake
of clarity in plotting, to share the same values of $\log g$ and $\log
T_{\rm eff}$, and to differ from each other only in the assumed
$\alpha$.

In order to have \smax\ in 1D models match that of 3D simulations,
the MLT parameter $\alpha$ must be selected (and varied with
evolution) so that the evolution track matches the locus of the 3D
simulations.  To illustrate this, we will consider a particular \logg\
and \lteff\ represented by the vertical dashed line in the left panel
of the figure.  The three example main-sequence MLT models (identified
on each evolution track with circles) do not match the {\smax}
predicted by 3D simulations, but it is clear that a value for $\alpha$
could be selected to reproduce the 3D simulated {\smax} in the 1D MLT
model.  The intersection of the vertical line with the locus of the 3D
simulations thus yields the correct entropy calibrated value of \smax\
for the model with this particular {\logg} and {\lteff}.  The
corresponding value of $\alpha$ that will result in the desired
{\smax} can then be read off the plot on the right hand panel of
Figure~\ref{fig:entropycal}.

As we emphasized earlier, a mapping between the entropy calibrated
$\alpha$ and \smax\ will not be general.  It depends sensitively
on various aspects of the stellar surface conditions, some of which
are imperfectly understood, and are treated differently by various
researchers.  The specific calibrated value of  $\alpha$ is thus model
dependent,  as it depends on the details of the inputs used in the
stellar evolution calculations.  {The calibration process described
above cannot be carried out once (for each chemical composition) to
determine a value of $\alpha$ that can applied to all other stellar
models.  Instead, the procedure illustrated in Fig.
\ref{fig:entropycal} would need to be applied as part of the stellar
evolution calculation.}  The calibration is also particularly
sensitive to the details of the $T$-$\tau$ relation. This effect is
well-known in solar model construction, where, for example a larger
value of $\alpha$ is needed to match the solar radius using a
Krishna-Swamy model atmosphere than an Eddington approximation model
atmosphere. {This is an important distinction between previous
attempts at mixing-length calibration, and the technique we present in
this letter.  For the calibration to remain generally applicable to
stellar models, it must relate to the thermal structure of the
convective envelope, and for the purpose of improving the accuracy of
stellar radii, calibrating against \smax\ is  appropriate.  Our method
shows that the evolution of \smax\ in the \logg\ - \lteff\ plane can
be presented in a convenient functional form, and we leave the final
step of mapping from \smax\ to $\alpha$ to the modeller. }

\section{Conclusion}

We have provided a simple prescription of how 3D simulations can be
used to calibrate the mixing length parameter in 1D stellar models.
The calibration procedure based on the {specific entropy adiabat}
presented in this paper provides a  reliable way, based on simple
and well established physical principles, of evaluating theoretically
stellar radii of late-type stars.  In this respect, the method is very
general, since it depends only on the chemical composition and the
well understood thermodynamic properties of deep convective
stellar envelopes.

\acknowledgements {We would like to thank the referee for comments that
have helped in improving this paper.}

\newpage

\end{document}